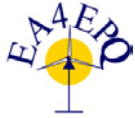
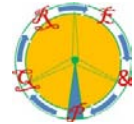

# PWM strategy with harmonics injection and modulated frequency triangular carrier. A review.

A. Ruiz-Gonzalez[1], M. Meco-Gutierrez[1], F.M. Perez-Hidalgo[1], F. Vargas Merino[1] and J. Heredia-Larrubia[2]

[1] Department of Electrical Engineering
E.T.S.I.I., Málaga University
Dr. Ortiz Ramos, S/N, 29071 Málaga (Spain)
Phone/Fax number: +0034 951 952353, e-mail: afruiz@uma.es, mjmeco@uma.es, fmperez@uma.es, fvagas@uma.es,

[2] Department of Technology Electronic
E.T.S.I.I., Málaga University
Dr. Ortiz Ramos, S/N, 29071 Málaga (Spain)
Phone/Fax number: +0034 951 952XXX, e-mail: jrheredia@uma.es

**Abstract.** A new, programmed pulse-width modulation (PWM) technique to control power inverters, which uses a harmonic-injection modulator and a frequency-modulated triangular carrier, synchronized with the modulating signal is presented in this paper. The instantaneous carrier frequency is adjusted according to a periodic function synchronized with the fundamental term of the modulating signal, in order to maintain the average value of the instantaneous frequency as an odd positive integer multiple of 3, for each period of the modulating signal—which is known as the average modulation order. The advantages of using the proposed technique over the conventional PWM techniques are the reduction in the total harmonic distortion and shift the frequency up of the temporal harmonics for any average modulation order. The experimental results show the viability of optimizing the time harmonics generated to minimize the vibrations in an induction motor or avoid the resonant frequencies.
The mathematical formulation for the output modulated voltage is defined and the results are also checked experimentally and compared to a sinusoidal PWM technique.

## 1. Introduction

The traditional carrier-based pulse width modulation technique compares a high frequency triangular carrier signal with a constant value, and a sinusoidal modulator signal. By comparing both, the modulated signal that turns on/off the power switches is obtained. Although this technique is easy to implement, its main disadvantage is that it generates a fundamental term with small amplitude and a significant amount of undesired harmonics.

Voltage source inverters are used for generating sinusoidal fundamental voltage can be either voltage or current controlled. From among the different controlled PWM techniques, a generated carrier-based PWM type can be used due to the inherent simplicity, if a fast dynamic response is not required. The output voltage can be controlled in several different ways. The most common method consists of controlling the pulse width applied to the electronic switches with carrier-based pulse width modulation (PWM) for two or multilevel inverters [1]. For the first type, the wave frequency spectrum obtained at the output contains many harmonics, which increase the amount of loss, vibration and noise, thereby reducing the performance of the system. With space vector based hybrid PWM techniques is possible to reduce the current ripple [2]. Other way could be using a hybrid method combining both the generated-carrier-based PWM and the space-vector-based hybrid PWM [3].

Researchers have developed multilevel techniques – instead of three levels- to reduce thermal stress, to reduce the acoustic noise or vibration level produced, to avoid resonance frequencies excitation or space harmonics for sensitive frequencies (with high winding factor) in the lowest vibratory modes, to reduce THD and DF factors, and to reduce the common-mode voltage [4]-[8]. Even so, this method presents a disadvantage due to the greater complexity for the inverter structure.

With a single inverter structure, when the inverter operates for a low number of pulses per period, M, the THD value is high. For this case, a PWM technique with carrier modulation frequency, HIPWM-FMTC [16]-[18] allows reducing the amplitude of the side band harmonics. The technique is used to decrease acoustic noise and to achieve a reduction in THD.

Efficient electrical and acoustic results were presented using the modulation technique called HIPWM-FMTC2 (harmonics injection PWM frequency modulated triangular carrier) that uses a linear modulation law instead of a sinusoidal modulation to change the carrier frequency [9]. Random PWM techniques are used to avoid mechanical resonances [10]-[13].

Other solution could be to use a PFM (Pulse Frequency Modulated). This strategy is based on a frequency modulation of the triangular carrier through a periodical signal [14]-[15].

It is possible to match its highest value (maximum frequency of the carrier signal) and cancel the inverter switching in the periods where the modulator signal has a reduced or null slope [19]-[21]. It will be a slope modulation strategy for generated PWM.

This paper focuses on a modified HIPWM-FMTC (Harmonics Injection PWM Frequency Modulated Triangular Carrier) technique. The present paper examines the results for the inverter output using a sinusoidal truncated function to control the triangular carrier signal. The main advantage of the new modulation function is that it permits to obtain output voltage through Bessel function. Consequently, any harmonic term can be annulled, adjusting the parameters for the carrier modulation. Likewise is possible to modify the electrical spectrum on a wider range using one parameter while the number of pulses per period remains unchanged. Consequently, any harmonic tem can be reduced or annulled by adjusting the carrier modulation parameters.

## 2. HIPWM-FMTC3 Strategy.

This technique, named HIPWM-FMTC3, obtains the commutation pulses by comparing two signals: a sinusoidal signal with harmonics injection wave, and a triangular carrier signal with variable frequency, as is showed in Fig. 1. The triangular carrier frequency is modulated by a periodical signal by adjusting the instantaneous modulation order (relationship between carrier and modulator signals). Using the strategy presented, the carrier frequency can be raised depending on k value, and therefore, reducing the harmonic terms of the side bands over the frequencies: A(1-k)·f –defined only for positive values- in the electrical spectrum, being f, the modulator signal frequency, (2).

With the original technique, HIPWM-FMTC [17-18], the instantaneous modulation order function is a sinusoidal function. Hence, using this HIPWM-FMTC strategy, a frequency $f_{max}$ with double value of f, is the maximum elongation possible. With the proposed technique, the elongation can be increased according the k value, following (2). The maximum instantaneous frequency for the carrier signal, $f_m$, is synchronized with a discontinuous sinusoidal signal for the null value of the modulator signal for a given M value (see Fig. 1).

The carrier wave increases the number of modulator signal samples during the interval where the modulator signal has the greatest slope and canceling these samples during the intervals when the modulating signal is closest to the maximum or minimum levels. The medium modulation order, $\bar{M}$, is fixed by the relation between A and k values. The instantaneous pulsation function that controls the carrier signal is:

$$\omega_i = \frac{d\theta}{dt} = A \cdot \left[\cos^2(\omega_m t) - k\right] \quad (1)$$

The relationship between A and k that it defines the modulation function for the carrier wave and it is obtained using:

$$\bar{M} = \frac{1}{\pi}\int_0^\pi A \cdot \left(\cos^2(\omega_m t) - k\right) d\omega t \quad (1)$$
$$= \frac{A}{\pi}\int_0^{\cos^{-1}\sqrt{k}} \left[\cos^2(\omega_m t) - k\right] \cdot d\omega t \quad (2)$$

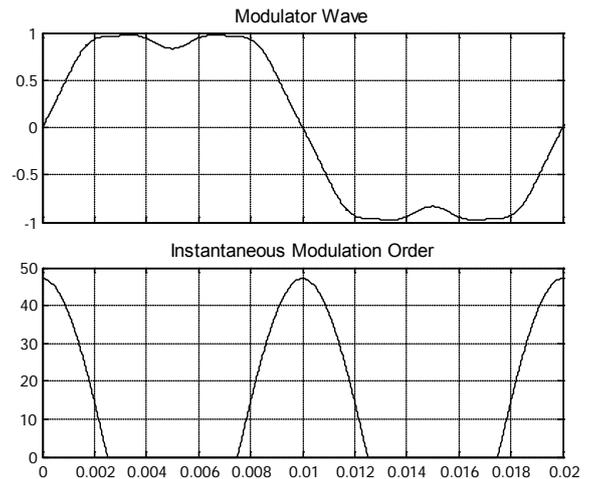

Fig. 1. Modulator Wave for 50 Hz and Instantaneous Modulation Order with HIPWM-MTC3 for k=0.5, A=94.2478$\omega_m$, M=15, a) Modulator signal, b) Instantaneous modulation order.

The instantaneus modulation order will be a discontinued function syncrhonized with the modutation signal, $\omega_m$, as it is showed on Fig. 2.

The area under instantaneous modulation order –Fig. 2 (b) will be $\bar{M}$. For each value, positive integer and odd number, infinite couples could be defined. The synchronization between modulator and instantaneous modulation order (function that defines the frequency evolution for carrier signal) signals ensures that the maximum value for the carrier frequency matching with the interval where the modulator signal presents the higher slope (for 0 and π rd with maximal IGBTs switching speed). The evolution follows a discontinuous quadratic sinusoidal function for the purpose to cancel the pulses around π/2 y 3π/2 phases.

The maximum carrier instantaneous frequency will be A·(1-k) times the modulator signal frequency, and it will be limited in each case by the IGBTs maximum switching speed. As k rises, the carrier frequency excursion is increased during a modulation signal period for a given $\bar{M}$ value. And therefore, the side band wide in the harmonic spectrum will be expanded. However, as these side bands wide grow, this terms have lower amplitude due to increasing n on the Bessel function.

It is clear from Fig. 2 that the converter produces switching pulses only during 4/10 part of the total period (20 ms for 50 Hz) with $\bar{M}$=15, k=0.5; A=94.2478$\omega_m$ (2).

The excursion for the carrier signal will be $A\cdot(1-k)\cdot\omega_m = 47.1239\cdot\omega_m$. The pulses are concentrated around $m\cdot T/2$ instant, being T the period for the series m=0,1,2…. The frequency is maximal for those time intervals; (94.2478/2) times the modulator frequency, instead of 15 times the modulator frequency with a classical technique SPWM.

Each k value allows a different A value when $\bar{M}$ is defined. If k is increased to 0.8, near to their upper limit, the excursion for the carrier frequency grows till 77.45 $\omega_m$; this is, 77.45/30 times, the excursion capability for HIPWM-FMTC [17].

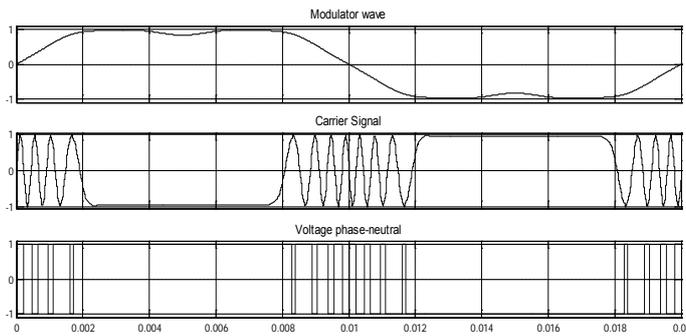

Fig. 2. HIPWM-FMTC3 technique (k=0.5 y A=94.2478$\omega_m$).

The output voltage fundamental and the harmonics terms will be:

$$\frac{a_0}{2} = \frac{1}{2\pi}\int_0^{2\pi} f(t)\cdot d(\omega_p t) = \frac{1}{2\pi}\left[\int_{-\pi}^{-\alpha} -\frac{E}{2}d(\omega_p t) + \int_{-\alpha}^{\alpha}\frac{E}{2}d(\omega_p t) + \int_{\alpha}^{\pi}-\frac{E}{2}d(\omega_p t)\right] = \ldots$$
$$= \frac{E}{2}\left(2\frac{\alpha}{\pi}-1\right)$$
(3)

$$a_n = \frac{1}{\pi}\int_0^{2\pi} f(t)\cdot\cos(n\omega_p t)\cdot d(\omega_p t) =$$
$$\frac{1}{\pi}\left[\int_{-\pi}^{-\alpha}-\frac{E}{2}\cdot\cos(n\omega_p t)\cdot d(\omega t) + \int_{-\alpha}^{\alpha}\frac{E}{2}\cdot\cos(n\omega_p t)\cdot d(\omega_p t) + \int_{\alpha}^{\pi}-\frac{E}{2}\cdot\cos(n\omega_p t)\cdot d(\omega_p t)\right] =$$
$$\ldots = \frac{4}{n\pi}\frac{E}{2}\sin(n\alpha)$$
(4)

$$b_n = \frac{1}{\pi}\int_0^{2\pi} f(t)\cdot\sin(n\omega_p t)\cdot d(\omega_p t) =$$
$$\frac{1}{\pi}\left[\int_{-\pi}^{-\alpha}-\frac{E}{2}\cdot\sin(n\omega_p t)\cdot d(\omega_p t) + \int_{-\alpha}^{\alpha}\frac{E}{2}\cdot\sin(n\omega_p t)\cdot d(\omega_p t) + \int_{\alpha}^{\pi}-\frac{E}{2}\cdot\sin(n\omega_p t)\cdot d(\omega_p t)\right] =$$
$$= 0$$
(5)

α value is obtained with the curves intersection:

$$1.15\cos(\omega_m t) - 0.27\cos(3\omega_m t) - 0.029\cos(9\omega_m t) = \frac{2\alpha}{\pi}-1 \Rightarrow$$ (6)
$$\alpha = \frac{\pi}{2} + \frac{\pi}{2}(1.15\cos(\omega_m t) - 0.27\cos(3\omega_m t) - 0.029\cos(9\omega_m t))$$

$$V_A(t) = \frac{a_0}{2} + \sum_{n=1}^{\infty} a_n\cdot\cos(n\omega_p t) + \sum_{n=1}^{\infty} b_n\cdot\sin(n\omega_p t) =$$
$$= \left(\frac{E}{2}\right)[1.15\cos(\omega_m t) - 0.27\cos(3\omega_m t) - 0.029\cos(9\omega_m t)] +$$
$$+ \left[\frac{4}{\pi}\left(\frac{E}{2}\right)\sum_{n=1}^{\infty}\left[\frac{1}{n}\sin\left(\frac{n\pi}{2}+\frac{n\pi}{2}H\right)\cdot\cos(n\omega_p t)\right]\right]$$
$$H = 1.15\cos(\omega_m t) - 0.27\cos(3\omega_m t) - 0.029\cos(9\omega_m t)$$
(7)

The first term is the fundamental for the modulator frequency, and the second term represents the harmonics terms for the carrier frequency, $\omega_p$. The instantaneous phase in (2) is:

$$\theta = \int\left[A\cdot(\cos^2(\omega_m t) - k)\right]dt =$$
$$\int\frac{A}{2}(1 + 2\cos(\omega_m t) - 2k)dt =$$
$$= A\cdot(0.5-k)\cdot t + \frac{A}{4\omega_m}\sin(2\omega_m t)$$
(8)

The phase-neutral voltage $V_{A0}$ will be:

$$V_A(t) = \frac{a_0}{2} + \sum_{n=1}^{\infty} a_n\cdot\cos(n\omega_p t) + \sum_{n=1}^{\infty} b_n\cdot\sin(n\omega_p t) =$$
$$\left(\frac{E}{2}\right)[1.15\cos(\omega_m t) - 0.27\cos(3\omega_m t) - 0.029\cos(9\omega_m t)] +$$
$$+ \left[\frac{4}{\pi}\left(\frac{E}{2}\right)\sum_{n=1}^{\infty}\begin{bmatrix}\frac{1}{n}\sin\left(\frac{n\pi}{2}+\frac{n\pi}{2}(1.15\cos(\omega_m t) - 0.27\cos(3\omega_m t) - 0.029\cos(9\omega_m t))\right)\cdot \\ \cdot\cos\left(n(A\cdot(0.5-k)\cdot t + \frac{A}{4\omega_m}\sin(2\omega_m t))\right)\end{bmatrix}\right]$$

$t_2 = \pi - \sin^{-1}\sqrt{k} > t > t_1 = \sin^{-1}\sqrt{k}$
$t_4 = 2\pi - \sin^{-1}\sqrt{k} > t > t_3 = \pi + \sin^{-1}\sqrt{k}$

$V_A(t) = 1 \longrightarrow t \leq t_1 = \sin^{-1}\sqrt{k}$
$V_A(t) = -1 \longrightarrow t_2 = \pi - \sin^{-1}\sqrt{k} \leq t \leq t_3 = \pi + \sin^{-1}\sqrt{k}$
$V_A(t) = 1 \longrightarrow t \geq t_4 = 2\pi - \sin^{-1}\sqrt{k}$
(9)

## 3. Results.

The result for the equation (9) is presented on fig. 3.

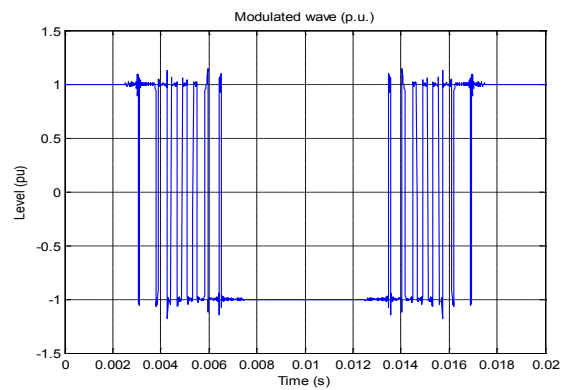

Fig. 3. Modulated wave with the technique proposed, $V_A$ (k=0.5, A=94.2478$\omega_m$, M=15).

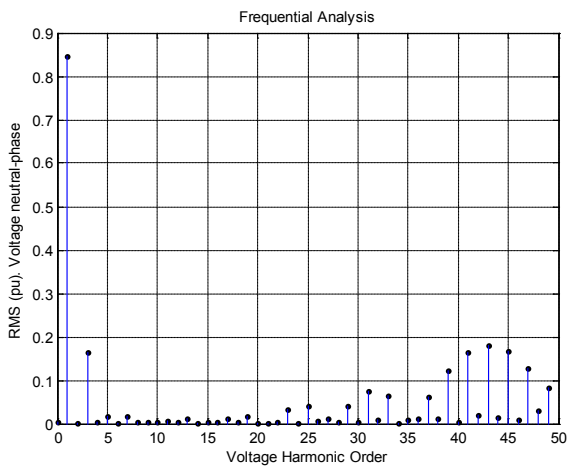

Fig. 4. Frequency spectrum for new proposed technique. Phase neutral voltage, $V_A$ (M=15, k=0.5, A=94.2478$\omega_m$).

## 4. Conclusions.

A new PWM modulation technique is presented. The mathematical expression for the phase signal $V_A$ at the output of the inverter was described. –Being a PWM technique with an FM carrier signal, each lateral band of the electric spectrum was divided into a greater number of significant harmonic terms. A control parameter is used to modify the electric spectrum of the output signal. However, the modulation order is maintained. The electrical spectrum could be modified, increasing the frequency for the resulting harmonic components, reducing the THD factor, with respect to a conventional technique for the same M value, and the terms of the lateral bands were also dispersed.